\documentclass{IEEEtran}
\usepackage{amssymb}
\usepackage{algorithmic}
\usepackage{amsfonts}
\usepackage{amsmath}
\usepackage{algorithm}
\usepackage{graphicx,subfigure,cite,multicol,multirow,diagbox,booktabs,array}

\usepackage[dvips]{color}
\usepackage{float}
\usepackage{bm}
\usepackage{footnote}
\usepackage{threeparttable}

\ifCLASSINFOpdf
\else
\fi
\begin{document}
\title{Beamforming and Transmit Power Design for Intelligent Reconfigurable Surface-aided Secure Spatial Modulation}

\author{Feng Shu,~Xinyi Jiang,~Wenlong Cai,~Weiping~Shi,~Mengxing Huang,\\~Jiangzhou Wang,~\emph{Fellow},~\emph{IEEE}, and Xiaohu You,~\emph{Fellow},~\emph{IEEE}
\thanks{This work was supported in part by the National Natural Science Foundation of China (Nos. 62071234, 61771244)(Corresponding authors: Feng Shu and MengXing Huang).}
\thanks{Feng Shu and Mengxing Huang are with the School of Information and Communication Engineering, Hainan Unversity, Haikou, 570228, China(e-mail:,shufeng0101@163.com).}
\thanks{Wenlong Cai is with the Second Academy of Aerospace Science and Industry, Beijing, 100089, China.}
\thanks{Xinyi~Jiang is with School of Electronic and Optical Engineering, Nanjing University of Science and Technology, Nanjing, 210094, China.}
\thanks{Jiangzhou Wang is with the School of Engineering and Digital Arts, University of Kent, Canterbury CT2 7NT, U.K. (E-mail: j.z.wang@kent.ac.uk)}
\thanks{Xiaohu You is  with the  School  of Information Science and Technology, Southeast University, Nanjing 210096, China(E-mail: xhyu@seu.edu.cn).}
}
\maketitle
\begin{abstract}
Intelligent reflecting surface (IRS) is a promising solution to build a programmable wireless environment
for future communication systems, in which the reflector elements steer the incident signal in fully customizable
ways by passive beamforming. In this paper, an IRS-aided secure spatial modulation (SM) is proposed, where the IRS
perform passive beamforming and information transfer simultaneously by adjusting the on-off states of the reflecting
elements.
We formulate an optimization problem to maximize the average secrecy rate (SR) by jointly optimizing the passive beamforming
at IRS and the transmit power at transmitter under the consideration that the direct pathes channels from transmitter to receivers
are obstructed by obstacles.
As the expression of SR is complex, we derive a newly fitting  expression (NASR) for the expression of traditional approximate SR (TASR),  which has simpler closed-form and more convenient for subsequent optimization.
Based on the above two fitting expressions, three beamforming methods, called maximizing NASR via successive convex approximation
(Max-NASR-SCA), maximizing NASR via dual ascent (Max-NASR-DA) and maximizing TASR via semi-definite relaxation (Max-TASR-SDR)
are proposed to improve the SR performance.
Additionally, two transmit power design (TPD) methods are proposed based on the above two approximate SR expressions, called
Max-NASR-TPD and Max-TASR-TPD.
Simulation results show that the proposed Max-NASR-DA and Max-NASR-SCA IRS beamformers harvest substantial
SR performance gains over Max-TASR-SDR. For TPD, the proposed Max-NASR-TPD  performs better than Max-TASR-TPD.
Particularly, the Max-NASR-TPD has a closed-form solution.

\end{abstract}

\begin{IEEEkeywords}
Spatial modulation, intelligent reflecting surface beamforming, physical layer security, secrecy rate,
transmit power design.
\end{IEEEkeywords}

\IEEEpeerreviewmaketitle

\section{Introduction}
AS a multiple-input multiple-output (MIMO) transmission scheme, spatial modulation (SM) exhibits a range of
advantages and it has been recognized as a promising transmission option for MIMO systems \cite{spatialSurvey}. The concept of
SM was first proposed in \cite{YA} whose main idea was to carry additive bit information via antenna indices \cite{RY2008}.
Different from the two typical forms of MIMO, Bell Laboratories Layer Space-Time (BLAST) \cite{Foschini2010Layered} and space time coding (STC) \cite{Yu2015Power}, SM may strike a good balance between spatial multiplexing and diversity and is called the third way between BLAST and STC. Compared to BLAST and STC, SM becomes more attractive
due to its advantages of no inter-channel interference (ICI) , inter-antenna synchronization (IAS) \cite{MR2011} and the use of less active radio frequency (RF) chains. Thus, it is also a green wireless transmission technique.

However, for such a SM system, due to the broadcast nature of wireless channel, like other MIMO systems, it is
very possible that the confidential messages are intercepted by unintended receivers. As how to achieve a secure
transmission is becoming a hot research topic in wireless networks, physical layer security \cite{XC2017,WT2015} in MIMO systems has
been widely investigated. There are several ways to improve the performance of SM including transmit antenna
selection \cite{shu2018two,Rajashekar2013Antenna,Xia2018AS,Xia2019AN,WY2020}, linear precoding \cite{Jin2015Linear,Y2011T}, power allocation \cite{shu2019high,xia2018power} and so on. In \cite{zhao2019}, the author
enhanced the legitimate security by jointly precoding optimization with and without eavesdroppong channel state
information (CSI). In \cite{wang2020,zhao2020},  the authors proposed a joint precoding optimization scheme which applies nonlinear
energy harvesting (EH) model. In \cite{WL2015,XY2018}, security was enhanced by emitting the artificial noise (AN) onto the
null-space of the desired channel and the latter derived the closed-form approximated expression of ergodic secrecy
rate (ESR) in perfect and imperfect CSI, respectively. In \cite{shu2021twc}, the authors proposed three precoding methods and
five transmit antenna subarray selection methods to improve the security performance under the hybrid SM systems.
Additionally, the authors in \cite{jxy2021} considered the malicious attacks from the eavesdropper and proposed several
effective beamforming methods to eliminate the interference from eavesdropper.

As a matter of fact, the intelligent reconfigurable surface (IRS) has emerged as a revolutionary
technology for improving the coverage and energy/spectrum efficiency of future wireless communications \cite{Wu2021IRS}.
Specifically, IRS consists of a large number of small, low-cost, and passive elements of only reflecting the incident
signal with an adjustable phase shift without complex precoding and radio frequency processing. Equipped
 with a smart controller, the IRS is able to intelligently adjust the phases of incident electromagnetic waves
 to increase the received signal energy, expand the coverage region, and alleviate interference, so as to enhance
 the communication quality of wireless networks.

There have been several innovative studies on the IRS-assisted wireless communication systems by jointly optimizing
the beamforming vector and the phase shifts at the IRS \cite{shi2021enhanced,whi2021,hong2020,pan2020,hua2021}. An IRS-aided secure wireless
information and power transfer (SWIPT) system was studied in \cite{shi2021enhanced}, and the authors adopted the SDR and alternating
optimization algorithm to maximize the harvested power. Additionally, multigroup and multicell MIMO communications
were studied in \cite{whi2021,hua2021}, respectively, where the former proposed to invoke an IRS at the cell boundary of multiple
cells to assist the downlink transmission to cell-edge users, and the latter jointly optimized thetransmit beamformer,
artificial noise(AN) vector and phase shifts at the IRS for minimizing the transmit power at Alice subject
to the secrecy rate constraints as well as the unit modulus constraints of IRS phase shifts. In \cite{hong2020}, the authors jointly
optimized the precoding matrix at the BS, covariance matrix of AN and phase shifts at the IRS for maximizing SR under
the consideration of an artificial noise(AN)-aided secure MIMO wireless communication system. \cite{hua2021} considered
the fairness among cell-edge users, and the authors  aimed at maximizing the minimum achievable
rate of cell-edge users by jointly optimizing the transmit beamforming at the BSs and the phase shifts
at the IRS.

As mentioned above, SM is a special MIMO technology of activating one transmit antenna with one transmit
antenna and exploits the index of the active antenna for information transfer. The undeniable potential
of both SM and IRS based communication schemes has attracted more and more attention nowadays. The concept of
IRS-assisted communications was first brought to the realm of SM in \cite{Basar}. In \cite{Yanwenjing}, the
authors applied SM principle to the IRS by adjusting the ON (active) and OFF (inactive) status of each reflecting
element. Therefore, the IRS can deliver additional information by adopting SM on the index of the reflecting
elements. In \cite{Basar}, the authors investigated the IRS-aided receive SM (RSM) technique.
Inspired by \cite{Basar}, the authors in \cite{Mateng} extended its structure to combine the transmit and
receive antenna indices for joint spatial modulation by shaping the reflecting beam with IRS. It is worth
mentioning that conventional SM cannot combine transmit SM (TSM) and RSM at the same time, due to the limitation
of a single activated transmit antenna.

Prior works on IRS-aided SM systems are mainly focused on maximizing the received signal strength, achievable rate,
spectral efficiency and outage probability which also showed that the IRS-aided SM system outperforms the conventional
SM system. However, there is no study on the SR performance of IRS-assisted SSM system, which might
be an potential way to make a significant improvement in SR performance. To investigate this issue, we propose an IRS-aided
secure SM (IRS-SSM) system. In this system, we activate a subset of the IRS elements for
reflecting a beam towards the intended destination, while exploiting the index combination of the ON-state IRS elements
to implicitly convey the spatial information of the IRS. Additionally, considered that the transmit channels from transmitter to receivers are blocked by obstacles, where the IRS is necessary for
communication. As there exists illegal receiver to eavesdrop the confidential information, we optimize
the beamforming at IRS and the transmit power at transmitter jointly to maximize SR. The main contributions of this
paper are summarized as follows:
\begin{enumerate}
\item  An IRS-aided secure SM system model is established, where the direct path channels in
the communication system from transmitter to receivers are obstructed by obstacles. Additionally, the transmitter is
equipped with single antenna, the desired receiver and eavesdropping receiver are equipped with multiple antennas.
Since the IRS elements have been divided into multiple subsets equally, the spatial bits are carried by activating
one of the subsets of IRS rather than the transmitter/receiver antenna, while the APM symbols are reflected by the
active IRS subset by adjusting the activated subset to ON state. Each IRS subset reflects single bit stream by using
multiple reflecting elements with secure beamforming. This will create spatial diversity and will be exploited to
improve the security performance of the IRS-SSM.
\item  To improve the secrecy performance, we first formulate the optimization problem with the aim of maximizing
secrecy rate subject to constrains of transmit power limit and unit modulus of IRS phase shifts. As the objective
function is different from the traditional SSM, we rederive it in Section II. Additionally, as the objective function
does not have the closed-form expression, which increases the difficult of subsequent optimization, we review
the traditional approximated secrecy rate (TASR) expression and propose a new approximated secrecy rate (NASR)
expression. Meanwhile, the NASR fits the secrecy rate curves well and has a simple closed-form expression, which facilitates the
further optimization work.
\item  To improve the secrecy performance, three IRS beamforming methods, based on the above two fitting expressions,
called maximizing NASR via successive convex approximation (Max-NASR-SCA), maximizing NASR via dual ascent
(Max-NASR-DA) and maximizing (TASR) via semi-definite relaxation (Max-TASR-SDR) are proposed to improve the
SR performance. Due to the fact that the NASR has  simpler expression than the TASR,  more effective algorithms  are proposed to approach the optimal solution better. Simulation results show that the SR performance
of the proposed Max-NASR-DA is better than proposed Max-TASR-SDR and proposed Max-NASR-SCA. In particular, the proposed
Max-NASR-SCA has a better SR performance than Max-TASR-SDR in the medium and high signal-to -noise (SNR) regions.
\item  In order to further improve the secrecy performance, two secure transmit power design (TPD) methods are proposed based on
the NASR expression and TASR expression, respectively. Max-NASR-TPD is proposed based on the NASR expression which
has the sum of ratio form. Hence, we first transform the objective function from fractions to  integrations by using
the quadratic transform method. Then, by solving the Karush-Kuhn-Tucker (KKT) conditions of Lagrange functions, we
have a closed-form solution towards the optimization problem with low complexity. For comparison, we propose Max-TASR-
TPD based on the TASR expression. As the expression is complex, we adopt the gradient ascent method to solve it. Simulation
results show that the proposed Max-NASR-TPD harvests a substantial SR performance gain over the Max-TASR-TPD.
\end{enumerate}

\emph{Notations:} Boldface lower case and upper case letters denote vectors and matrices, respectively. $(\cdot)^{H}$ denotes the conjugate transpose operation.  $\mathbb{E}\{\cdot\}$ represents expectation operation. $\|\cdot\|$ denotes 2-norm. $\hat{[\;]}$ represents the estimation operation. $\textbf{A}'$ represents a matrix that is different from the original matrix $\textbf{A}$ but has a linear transformation relationship with the original matrix.

\section{System Model}
\subsection{IRS-Aided Secure Spatial Modulation System}
\begin{figure*}[t]
\centerline{\includegraphics[width=0.75\textwidth]{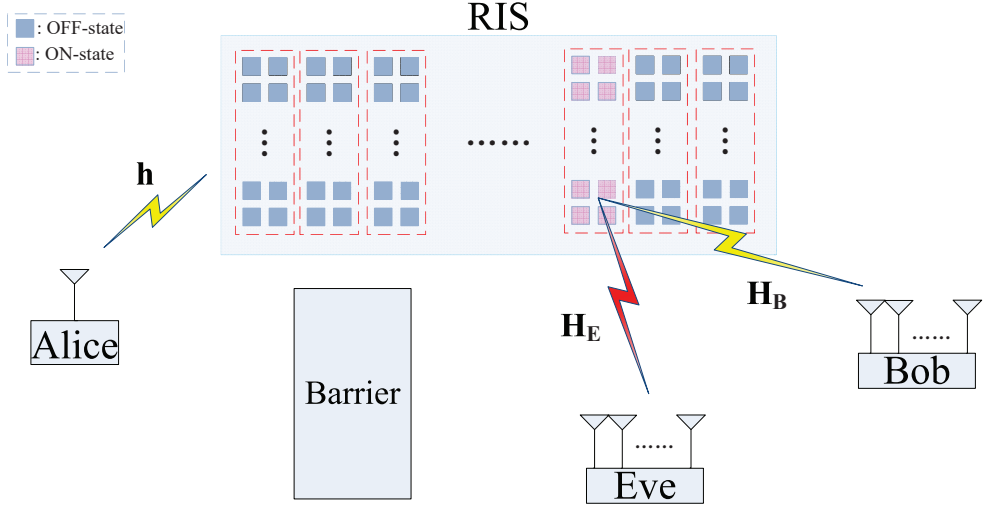}}
\caption{System model for secure Intelligent Reflecting Surface Aided Secure Spatial Modulation.}
\label{fig1}
\end{figure*}
As shown in Fig.~1, we consider a system where the transmitter (Alice), the legal receiver (Bob) and the eavesdropper (Eve) are equipped with  single antenna, $N_b$ and $N_e$ antennas respectively. We further assume that the IRS consists of $N$ low-cost passive reflecting elements. Meanwhile, we follow the hypothesis in \cite{Basar} that the direct link between the transmitter and receiver is obstructed. In the following, the IRS reflects signal only one time.

Different from the traditional spatial modulation system, where the system adds spatial module to the transmitter or the receiver, the system in this paper adds spatial module to the IRS by adjusting the ON-OFF states of the reflecting elements. We adopt the IRS-elements grouping method, where a total number of $N$ unit cell elements are divided into $G$ groups, each of which consists of $\bar{N}=N/G$ adjacent elements and the reflection coefficient of each element in the same group is different. Specifically, for each symbol duration, one of $G$ groups is randomly turned ON for reflecting the incident signals, and the remaining $(G-1)$ groups are deliberately turned OFF for realizing the SM scheme.

Accordingly, the transmission information is divided into two parts: the first one, denoted as $j\in \left\{1,2,\cdots,M\right\}$, is modulated as an $\mathcal{M}$-ary quadrature amplitude modulation (QAM) or phase shifted keying (PSK) symbol, and the second one, $i\in\left\{1,2,\cdots,G\right\}$ is used to activate the IRS subset $\mathcal S$, respectively. Therefore, the data rate of IRS-aided SM is expressed as
\begin{align}
R=\log_2(M)+\log_2(G).
\end{align}

Additionally, the reflection coefficient vector of IRS is denoted as $\bm{\theta}_{N\times1}\triangleq\left[\theta_1,\theta_2,\cdots,\theta_N\right]^T$, where $\theta_n=\beta_ne^{-j\phi_n}$, in which $\beta_n\in\left[0,1\right]$ and $\phi_n\in(0,2\pi]$ represent the common reflection amplitude and phase shift for the $n$-th reflection element. To ease the hardware design \cite{Basar}, the reflection amplitudes of the ON-state groups are set to the maximum value, i.e., $\beta_n=1$. Meanwhile, let $\textbf{h}_{t}=[\textbf{t}_1, \textbf{t}_2, \cdots, \textbf{t}_{G}]\in\mathbb{C}^{1\times N }$, $\textbf{H}_{B}=[\textbf{B}_1, \textbf{B}_2, \cdots, \textbf{B}_{G}]\in\mathbb{C}^{N_b\times N }$, and $\textbf{H}_{E}=[\textbf{E}_1, \textbf{E}_2, \cdots, \textbf{E}_{G}]\in\mathbb{C}^{N_e\times N }$ denote the channels from Alice to IRS, from IRS to Bob, and from IRS to Eve, respectively, where $\textbf{t}_g\in\mathbb{C}^{1\times \bar{N}}$, $\textbf{B}_g\in\mathbb{C}^{N_b\times \bar{N}}$ and $\textbf{E}_g\in\mathbb{C}^{N_e\times \bar{N}}$ denote the baseband element-wise channels from Alice to group g and from group g to
Bob and Eve, respectively. Without loss of generality, we assume that $\textbf{h}_t$, $\textbf{H}_B$ and $\textbf{H}_E$ are Rayleigh flat fading channels with each element of them obeying the independent identically distributed (iid) Gaussian distribution $\mathcal{CN}(0,1)$.

Moreover, due to the severe path loss
and high attenuation, the signals reflected by the IRS more than once have negligible power and hence can be ignored. Accordingly, the received signal at Bob and Eve are given by
\begin{align} \label{rey_b}
\textbf{y}_b=P_t\textbf{H}_{B}\rm{diag}\left\{ \bm{\theta} \right\}\rm{diag}\left\{\textbf{s}_i\right\}\textbf{h}_{t}\it{b_j}+\textbf{n}_{B},
\end{align}
\begin{align} \label{rey_e}
\textbf{y}_e=P_t\textbf{H}_{E}\rm{diag}\left\{ \bm{\theta} \right\}\rm{diag}\left\{\textbf{s}_i\right\}\textbf{h}_t\it{b_j}+\textbf{n}_{E},
\end{align}
, respectively, where $\textbf{s}_i=[\textbf{0},\cdots,\textbf{0},\textbf{e}_i,\textbf{0},\cdots,\textbf{0}]^T\in\mathbb{R}^{N\times 1}$, $\textbf{e}_i=[1,1,\cdots,1]^T\in\mathbb{R}^{\bar{N}\times1}$, $b_j$ is the digital symbol chosen from the $\mathcal{M}$-ary constellation for $j\in\mathbb{M}=\left\{1,2,\cdots,M\right\}$, and satisfies $\mathbb{E}|b_j|^2=1$,  $P_t=\beta^2P_s$ is the transmit power with the constraint of $P_t\leq N_t$, $0\leq\beta\leq1$ is the transmit power factor and $P_s=N_t$ is the total transmit power at Alice. $\textbf{n}_B\sim\mathcal{CN}(0,\sigma_b^2\textbf{I}_{N_b})$ and $\textbf{n}_E\sim\mathcal{CN}(0,\sigma_e^2\textbf{I}_{N_e})$ denote the complex additive white Gaussian noise (AWGN) vectors at Bob and Eve, respectively.

By changing the form $\rm{diag}\left\{ \bm{\theta} \right\}\rm{diag}\left\{\textbf{s}_i\right\}\textbf{h}_t$ as $\rm{diag}\left\{ \textbf{h}_t \right\} \bm\Phi \textbf{s}_i$, where $\bm\Phi=\rm{diag}\left\{\bm{\theta}\right\}$,
we can rewrite (\ref{rey_b}) and (\ref{rey_e}) as follows
\begin{align}
\textbf{y}_b=\beta^2P_s\textbf{H}_{B}'\bm{\Phi}\textbf{s}_i\it{b_j}+\textbf{n}_{B},
\end{align}
\begin{align}
\textbf{y}_e=\beta^2P_s\textbf{H}_{E}'\bm{\Phi}\textbf{s}_i\it{b_j}+\textbf{n}_{E},
\end{align}
where $\textbf{H}_{B}'=\textbf{H}_{B}\rm{diag}\left\{ \textbf{h}_t \right\}$ and $\textbf{H}_{E}'=\textbf{H}_{E}\rm{diag}\left\{ \textbf{h}_t \right\}$.
\subsection{Problem Formulation}
Here, we characterize the security by evaluating average SR, formulated as
\begin{align}
\bar{R_s}\!=\!\mathbb{E}_{\textbf{H}_B',\textbf{H}_E'}\!\!\left( \!\left[ I(b_j,\textbf{s}_i;\textbf{y}_b|\textbf{H}_B',\textbf{H}_E')\!-\!I(b_j,\textbf{s}_i;\textbf{y}_e|\textbf{H}_B',\textbf{H}_E') \right]\!^+ \!\right),
\end{align}
where $\left[ a \right]^+=\max\left\{ a,0 \right\}$ and $I(b_j,\textbf{s}_i;\textbf{y}_b|\textbf{H}_B',\textbf{H}_E')$ $I(b_j,\textbf{s}_i;\textbf{y}_e|\textbf{H}_B',\textbf{H}_E')$ are the mutual information over desired and eavesdropping channels with the finite and discrete complex signal set, respectively. Similar to \cite{wangli2015}, it can be derived as follows
\begin{align}\nonumber\label{Ib}
&I(b_j,\textbf{s}_i;\textbf{y}_b|\textbf{H}_B',\textbf{H}_E')\\&\nonumber=\!\!\int\!\!\sum_{i=1}^G\sum_{j=1}^Mp(\textbf{y}_b,\textbf{s}_i,b_j)\log_2\frac{p(\textbf{y}_b,\textbf{s}_i,b_j)}{p(\textbf{y}_b)p(\textbf{s}_i,b_j)}d\textbf{y}_b
\\&\nonumber=\frac{1}{GM}\sum_{i=1}^G\sum_{j=1}^Mp(\textbf{y}_b|\textbf{s}_i,b_j)\log_2\frac{GMp(\textbf{y}_b|\textbf{s}_i,b_j)}{\sum_{i'=1}^G\sum_{j'=1}^Mp(\textbf{y}_b|\textbf{s}_i,b_j)}d\textbf{y}_b
\\&\nonumber=\log_2{GM}-\frac{1}{GM}\sum_{i=1}^G\sum_{j=1}^M\mathbb{E}_{\textbf{n}_B}\left \{ \log_2\sum_{i'=1}^G\sum_{j'=1}^M \right .\\& \left .~~~\exp\left[ \frac{-\|\beta^2P_s\textbf{H}_{B}'\bm\Phi\left( \textbf{s}_ib_j-\textbf{s}_{i'}b_{j'} \right)+\textbf{n}_B\|^2+\|\textbf{n}_B\|^2}{\sigma_b^2} \right]  \right \}.
\end{align}
Similarly, we can derive the mutual information over eavesdropping channel as follows
\begin{align}\nonumber\label{Ie}
&I(b_j,\textbf{s}_i;\textbf{y}_e|\textbf{H}_B',\textbf{H}_E')\\&\nonumber=\log_2{GM}-\frac{1}{GM}\sum_{i=1}^G\sum_{j=1}^M\mathbb{E}_{\textbf{n}_E}\left \{ \log_2\sum_{i'=1}^G\sum_{j'=1}^M \right .\\& \left .~~~\exp\left[ \frac{-\|\beta^2P_s\textbf{H}_{E}'\bm\Phi\left( \textbf{s}_ib_j-\textbf{s}_{i'}b_{j'} \right)+\textbf{n}_E\|^2+\|\textbf{n}_E\|^2}{\sigma_e^2} \right]  \right \}.
\end{align}

Finally, our objective is to maximize the SR by designing the beamforming at IRS and the transmit power at Alice, which is casted as the following optimization problem
\begin{align}
\left( \textbf{P1} \right):~~&\max ~~  ~~~~R_s\\&\nonumber \textrm{subject} \ \textrm{to} \ |\bm{\Phi_{n,n}}|=1,
\\&~~~~~~~~~~~~\beta^2\leq1.
\end{align}


\section{Approximation of the Ergodic Mutual Information}
\subsection{Traditional Approximate Secrecy Rate Expression}
In accordance with the definition of the cut-off rate for traditional MIMO systems in \cite{Yu2016}, we have the cut-off rate for Bob and Eve as follows
\begin{align} \nonumber
\!I_0^{B}\!=&2\textrm{log}_2GM- \\
&\!\log_2\!\sum \limits_{i=1}^{GM}\!\sum \limits_{j=1}^{GM}\!\exp\!\left(\! \frac{\!-\! \beta^2P_s\textbf{d}_{ij}^H\bm\Phi^H\textbf{H}_B'^H\boldsymbol{\Omega}_{\rm{B}}^{-1}\textbf{H}_B'\bm\Phi\textbf{d}_{ij}}{4} \!\right)\!,
\end{align}
which can be derived similarly to Appendix A in \cite{Yu2016} with a slight modification as $\textbf{d}_{i,j}=\textbf{x}_i-\textbf{x}_j$ and $\textbf{x}=\textbf{s}_ib_j$. Similarly, the cut-off rate for Eve is given by
\begin{align} \nonumber
\!I_0^{E}\!=&2\textrm{log}_2GM- \\
&\!\log_2\!\sum \limits_{i=1}^{GM}\!\sum \limits_{j=1}^{GM}\!\exp\!\left( \! \frac{\!-\! \beta^2P_s\textbf{d}_{ij}^H\bm\Phi^H\textbf{H}_E'^H\boldsymbol{\Omega}_{\rm{E}}^{-1}\textbf{H}_E'\bm\Phi\textbf{d}_{ij}}{4} \!\right)\!,
\end{align}
where $\boldsymbol{\Omega}_{\rm{B}}=\sigma_b^2\textbf{I}_{N_b}$ and $\boldsymbol{\Omega}_{\rm{E}}=\sigma_e^2\textbf{I}_{N_e}$. Therefore, the TASR can be expressed as
\begin{align}
R_s^{a}=I_0^{B}-I_0^{E}.
\end{align}
\subsection{Proposed Newly Approximate Secrecy Rate Expression}
As previously stated, (\ref{Ib}) and (\ref{Ie}) can be applied to evaluate the
ergodic mutual information (EMI). With this
equation, we know that the mutual information is decided by the four variables, $\textbf{s}_i$, $b_j$, $\textbf{H}$ and $\textbf{n}$, and its expression is complex and does not have a closed-form expression. Authors in \cite{Ouyang2020,OY2020} tried to replace the above four variables with one variable SNR and they used the 1stOpt software to get a fitting expression which is simpler and has the closed-form expression. Here, we first try to fit the EMI in the same way as the authors in \cite{Ouyang2020}. However, we found that if we choose the expression of the same independent variable as \cite{Ouyang2020,OY2020}, the secrecy rate does not fit well. Therefore, we try to find another expression of independent variable which is derived from (8) and (9) directly.

Inspired by the formula given in [6, eq.~(4.3.34)], (11) can be rewritten as
\begin{align} \label{r_b}
\!\!I_0^{B}\!\!= \!\!-\!\log_2\!\!\sum \limits_{i=1}^{GM}\!\sum \limits_{j=1}^{GM}\!\frac{1}{(N_tM)^2}\!\!\int\! \!p(\textbf{y}_b|\textbf{x}_i)^{1/2}  p(\textbf{y}_b|\textbf{x}_j)^{1/2} d\textbf{y}_b,\!\!
\end{align}
where $\textbf{x}_{i}=\textbf{s}_{i}{b}_{j}$. For a given  channel $\textbf{H}$, assuming the decoded received signal $\textbf{y}_b$ is a complex Gaussian distribution, the corresponding conditional probability is
\begin{align} \label{pB}
p(\textbf{y}'_b|\textbf{x}_i)=\frac{1}{(\pi\sigma_b^2)^{N_b}}\exp\left( \|(\textbf{y}'_b-\beta^2P_s\textbf{H}_B\bm\Phi\textbf{x}_i)\|^2 \right).
\end{align}
By plugging $p(\textbf{y}'_b|\textbf{x}_i)$ and $p(\textbf{y}'_b|\textbf{x}_j)$ into (\ref{r_b}), we get
\begin{align}\label{r_b1}
\!\!I_0^{B}\!\!=&2\textrm{log}_2GM\nonumber\\&\!\!-\textrm{log}_2\sum \limits_{j=1}^{GM}\sum \limits_{j=1}^{GM}\int\!\left[\!\frac{1}{(\pi\sigma_b^2)^{N_r}}\!\exp\!\left( \!\|(\textbf{y}'_b-\beta^2P_s\textbf{H}_B\bm\Phi\textbf{x}_i)\|^2 \!\right) \right]^{\frac{1}{2}}\nonumber\\&\times\left[\!\frac{1}{(\pi\sigma_b^2)^{N_r}}\!\exp\!\left( \!\|(\textbf{y}'_b-\beta^2P_s\textbf{H}_B\bm\Phi\textbf{x}_j)\|^2 \!\right) \right]^{\frac{1}{2}} d\textbf{y}_b,
\end{align}
where the integrand can be simplified as
\begin{align}
\textbf{I}_1\!=&\frac{1}{(\pi\sigma_b^2)^{N_b}}\!\!\exp\!\!\left(\!\!-\|\textbf{y}'_b\|^2\!\!-\!\frac{1}{2}\|\beta^2P_s
\textbf{H}_B\bm\Phi\textbf{x}_i\|^2\!-\!\frac{1}{2}\|\beta^2P_s\textbf{H}_B\bm\Phi\textbf{x}_j\|^2\right.\nonumber\\&\left.
+R_e\left\{\left(\beta^2P_s\textbf{H}_B\bm\Phi\textbf{x}_i\right)*\textbf{y}'_b\right\}+R_e\left\{\left(\beta^2P_s\textbf{H}_B\bm\Phi\textbf{x}_j\right)*\textbf{y}'_b\right\}\right).
\end{align}
Then, substituting $\textbf{I}_1$ into (\ref{r_b1}) yields
\begin{align}
\!\!I_0^{B}\!\!=&2\textrm{log}_2GM-\textrm{log}_2\sum \limits_{j=1}^{GM}\sum \limits_{j=1}^{GM}\exp\left(-\frac{\|\textbf{H}_B\bm\Phi\left(\frac{\textbf{x}_i-\textbf{x}_j}{2}\right)\|^2}{\sigma_b^2}\right)
\nonumber\\&\nonumber\!-\!\textrm{log}_2\!\sum \limits_{j=1}^{GM}\!\sum \limits_{j=1}^{GM}\int\left[\exp\!\left(\!-\frac{\|\textbf{y}_b'\!-\!\textbf{H}_B\bm\Phi\left(\frac{\textbf{x}_i+\textbf{x}_j}{2}\right)\|^2}{\sigma^2}\!\right)\right.\\&\left.\times\frac{1}{(\pi\sigma_b^2)^{N_b}}\!\right]\!d\textbf{y}_b',
\end{align}
where the third item integral is equal to 1 since the integrand is a multi-variate Gaussian probability density function. Moreover, we can simplify the second item by the Jenson's inequality as follows
\begin{align}
\!\!I_0^{B}(\gamma)\!\!=&2\textrm{log}_2GM\!-\!\textrm{log}_2\exp\!\left(\!-\!\underbrace{\sum \limits_{j=1}^{GM}\sum \limits_{j=1}^{GM}\frac{\|\textbf{H}_B\bm\Phi(\textbf{x}_i-\textbf{x}_j)\|^2}{4\sigma^2} }_{\gamma} \right).
\end{align}

For the sake of improving the accuracy of the fitting expression, we choose $\gamma$ in the above as the independent variable. We plot the mutual information with respect to SNR for various values of G in Fig.~2, and we find that the sum of ratio term(s) curve fitting based expression
can be applied to accurately fit $I(b_j,\textbf{s}_i;\textbf{y}_b|\textbf{H}_B',\textbf{H}_E')$, which is expressed by
\begin{align}\nonumber\label{I_similar}
&~~~~I(b_j,\textbf{s}_i;\textbf{y}_b|\textbf{H}_B',\textbf{H}_E') \\&\approx \hat{I}_B^{(M,G)}(\gamma_b)=\sum_{i=1}^{k_G}\frac{\zeta_i^{(G)}\gamma_b}{\xi_i^{(G)}+\gamma_b}.
\end{align}

By using the open-source fitting software package 1stOpt, the fitting parameters $k_M$, $k_G$, $\left\{\zeta_i^{(M)}\right\}$ and $\left\{\xi_j^{(G)}\right\}$ can be found, which are listed in Table I.
\begin{table*}[htbp]
\centering
\caption{Fitting coefficients for M-PSK with different G in (\ref{I_similar})}
\begin{tabular}{ccccccccccc}
\hline
\hline
$M$                & $G$  & $k_M$ & $k_G$ & $\zeta_1^{(M)}$ & $\zeta_2^{(M)}$ & $\zeta_3^{(M)}$ & $\xi_1^{(G)}$& $\xi_2^{(G)}$& $\xi_3^{(G)}$ &RMSE\\
\hline
\hline
\multirow{4}{*}{8} &16 &\multirow{4}{*}{2} &\multirow{2}{*}{4} &2.007602 &-1.83073 &4.78618	&16.48645	&1.90012	&11.89605  &1.298E-3\\
&8 & & &3.22453 &4.62917 &-0.85445 &2.89612	&13.98985	&88.82823 &2.7413E-3\\
&4 & &\multirow{2}{*}{3} &4.67308  &-1.41513 &2.74124	&11.89605	&45.76453 &2.55830 &2.4591E-4\\
&2 & & &-61.65283	&2.007602	&64.640742	&16.48645	&1.90012 &15.62086 &4.1167E-4\\
\hline
\multirow{4}{*}{4} &16 &\multirow{4}{*}{2} &\multirow{2}{*}{4} &-69.79872 &73.46507 &2.33084	&14.87082	&14.01141	&2.57772  &1.298E-3\\
&8 & & &1.96523 &42.82407 &-39.79035 &1.13629	&5.11235	&5.57396 &1.081E-3\\
&4 & &\multirow{2}{*}{3} &-24.41596  &27.36585 &1.05006	&3.74558	&3.30759 &0.84914 &8.808E-4\\
&2 & & &0.044593	&-1.83073	&4.78618	&0.247079	&4.14794 &1.57405 &6.613E-4\\
\hline
\multirow{4}{*}{2} &16 &\multirow{2}{*}{2} &\multirow{2}{*}{4} &62.5582	&-58.9554	&1.39777	&9.93359	&10.61559	&1.97674 &9.612E-4\\
&8 & & &0.37940	&60.51929	&-56.90064	&1.02712	&6.658413	&7.10888 &8.234E-4\\
&4 &\multirow{2}{*}{1} &\multirow{2}{*}{3} &0.08038	&58.8141	&-55.8964   &0.39127	&4.63607 &4.89234 &8.312E-4\\
&2 & & &14.6211	&-15.6089	&2.9887  &13.1342	&12.6401 &1.9804 &9.087E-4\\
\hline
\hline
\end{tabular}
\begin{tablenotes}
\footnotesize
\item[*]~~~~~~~~~~~~~~~~~~~~
{$M$ represents the modulation order of M-PSK. RMSE denotes the root mean square error.}\\
\end{tablenotes}
\end{table*}Notice that the mutual information will tend to  $\log_2GM$ or 0 when $\gamma$ tends to $+\infty$ or $-\infty$.  In this table, RMSE is short for means root mean square error, standing for the gap between
the exact and approximated value.

To further investigate the
precision of (\ref{I_similar}), we compare the approximated and exact mutual information in Fig.~2, which is shown in the simulation part for different modulation schemes and different number of $G$. It can be seen from the figure that the approximation results are very close to the simulation results,
which verifies the precision of (\ref{I_similar}).

Similarly, we can get the approximate expression of the mutual information over wiretap channel as follows
\begin{align}\nonumber
&~~~~I(b_j,\textbf{s}_i;\textbf{y}_e|\textbf{H}_B',\textbf{H}_E') \\&\approx \hat{I}_E^{(M,G)}(\gamma_e)=\sum_{i=1}^{k_G}\frac{\zeta_i^{(G)}\gamma_e}{\xi_i^{(G)}+\gamma_e}.
\end{align}

Therefore, the optimization problem (\textbf{P1}) can be rewritten as follows
\begin{align}
\left( \textbf{P2} \right):&\max_{\beta,\bm\Phi} ~~~~~~\sum_{i=1}^{k_G}\frac{\zeta_i^{(G)}\gamma_b}{\xi_i^{(G)}+\gamma_b}-\sum_{i=1}^{k_G}\frac{\zeta_i^{(G)}\gamma_e}{\xi_i^{(G)}+\gamma_e}\\&\nonumber \textrm{subject} \ \textrm{to} \ |\bm{\Phi_{n,n}}|=1,
\\&\nonumber~~~~~~~~~~~~\beta^2\leq1,
\end{align}
where the expressions of $\gamma_b$ and $\gamma_e$ can be further simplified from the original form in (19) as follows
\begin{align}
\gamma_b&=\sum \limits_{i=1}^{GM}\sum \limits_{j=1}^{GM}\frac{\|\textbf{H}_B'\bm\Phi(\textbf{x}_i-\textbf{x}_j)\|^2}{4\sigma^2}\\&=\sum \limits_{j=1}^{GM}\sum \limits_{j=1}^{GM}\frac{\textrm{tr}\left(\textbf{H}_B'{\bm\Phi}\textbf{d}_{i,j}\textbf{d}_{i,j}^H\bm\Phi^H\textbf{H}_B'^H\right)}{4\sigma^2}
\\&=\frac{1}{4\sigma^2}\textrm{tr}\left(\sum \limits_{j=1}^{GM}\sum \limits_{j=1}^{GM}\textbf{H}_B'{\bm\Phi}\textbf{d}_{i,j}\textbf{d}_{i,j}^H\bm\Phi^H\textbf{H}_B'^H\right)
\\&=\frac{1}{4\sigma^2}\textrm{tr}\left(\textbf{H}_B'{\bm\Phi}\textbf{D}_{ij}\bm\Phi^H\textbf{H}_B'^H\right)
\\&=\frac{1}{4\sigma^2}\textrm{tr}\left(\textbf{D}_{ij}\bm\Phi^H\textbf{H}_B'^H\textbf{H}_B'{\bm\Phi}\right).
\end{align}
The derivation from (23) to (24) is achieved by denoting $\textbf{d}_{i,j}=\textbf{x}_i-\textbf{x}_j$, and the derivation from (24) to (25) and (26) to (27) is achieved by utilizing the trace property, i.e., $\textrm{tr}(\textbf{A}+\textbf{B}) = \textrm{tr}(\textbf{A})+\textrm{tr}(\textbf{B})$ and $\textrm{tr}(\textbf{A}\textbf{B}\textbf{C}) = \textrm{tr}(\textbf{BCA})$, respectively. Similarly, we can get the expression of $\gamma_e$ as follows
\begin{align}
\gamma_e=\frac{\textrm{tr}(\textbf{D}_{ij}\bm\Phi^H\textbf{H}_{E}'^H\textbf{H}_{E}'\bm\Phi)}{G\sigma_e^2}.
\end{align}

$\textbf{P2}$ is difficult to solve due to the non-concave objective function as well as the coupled optimization variables. However, we observe that the resultant problems can be efficiently solved when one of $\bm\Phi$ and $P_t$ is fixed. This thus motivates us to propose an alternating optimization based algorithm to solve $\textbf{P2}$ sub-optimally, by iteratively optimizing $\bm\Phi$ ($\textbf{P2-1}$) and $P_t$ ($\textbf{P2-2}$) with the other being fixed at each iteration until convergence is reached, as detailed in the next sections.
\section{Beamforming Design for given transmit power based on Approximate expression of SR}
For IRS-aided SM systems, the design of beamformer at IRS is necessary to improve the system performance. In this section, two beamformers at IRS, called Max-NASR-SCA and Max-NASR-DA, are proposed based on the proposed NASR to enhance the security of IRS-aided SM systems. Additionally, the Max-TASR-SDR based on the traditional ASR is proposed and used as a performance reference.
\subsection{Proposed Max-NASR-SCA}
Observing the optimization problem in ($\textbf{P2}$), we find it is the fractional programming problem actually. However, the conventional FP techniques mostly can only deal with the single-ratio or the max-min-ratio case rather than the multiple-ratio FP problems like ($\textbf{P2}$). Thus we propose the Max-NASR-SCA method which first decouples the numerator and the denominator of each ratio term, and then utilizes the SCA method based on SDR to solve the problem. First we rewrite optimization problems with $P_t$ fixed as follows
\begin{align}
\left( \textbf{P2-1} \right):&\min_{\bm\Phi} ~\sum_{i=1}^{k_G}\frac{\zeta_i^{(G)}P_t\textrm{tr}(\textbf{D}_{ij}\bm\Phi^H\textbf{H}_{E}'^H\textbf{H}_{E}'\bm\Phi)}{4\xi_i^{(G)}\sigma_e^2
+P_t\textrm{tr}(\textbf{D}_{ij}\bm\Phi^H\textbf{H}_{E}'^H\textbf{H}_{E}'\bm\Phi)}
\\&\nonumber~~~~~~-\sum_{j=1}^{k_G}\frac{\zeta_j^{(G)}P_t\textrm{tr}(\textbf{D}_{ij}\bm\Phi^H\textbf{H}_{B}'^H\textbf{H}_{B}'\bm\Phi)}{4\xi_j^{(G)}\sigma_b^2
+P_t\textrm{tr}(\textbf{D}_{ij}\bm\Phi^H\textbf{H}_{B}'^H\textbf{H}_{B}'\bm\Phi)}\\&\nonumber ~\textrm{s.} \ \textrm{t.} \ ~|\bm{\Phi}_{n,n}|=1.
\end{align}
It is easy to see that the problem $\textbf{P2-1}$ is equivalent to the following problem
\begin{align}
\left( \textbf{P2-1-1} \right):&\min_{\bm\Phi} ~~~~~~~\sum_{i=1}^{2k_G}\alpha_i\\&  \nonumber
 ~\textrm{s.} \ \textrm{t.} ~~~~~~~~\frac{h_i(\bm\Phi)}{g_i(\bm\Phi)}\leq\alpha_i, i=1,\cdots,2k_G\\& \nonumber
~~~~~~~~~~~~\ |\bm{\Phi}_{n,n}|=1,
\end{align}
where $\bm\alpha$ refers to a collection of variables $\left\{ \alpha_1,\cdots,\alpha_{2k_G} \right\}$, and when $\bm\Phi$ is fixed, the optimal $\alpha_i$ can be found in closed form expression as
\begin{align}
\alpha_i^*=\frac{h_i(\bm\Phi)}{g_i(\bm\Phi)},~~\forall i=1,\cdots,2k_G
\end{align}
where
\begin{align}
h_i(\bm\Phi)=U_i P_t\textrm{tr}(\textbf{D}_{ij}\bm\Phi^H\textbf{H}_{g}'^H\textbf{H}_{g}'\bm\Phi),
\end{align}
and
\begin{align}
g_i(\bm\Phi)=4Q_i\sigma_g^2+P_t\textrm{tr}(\textbf{D}_{ij}\bm\Phi^H\textbf{H}_{g}'^H\textbf{H}_{g}'\bm\Phi),
\end{align}
where g stands for B (Bob) or E (Eve), $U_i \in \left\{ \zeta_i^{(G)}, -\zeta_j^{(G)}\right\}$, $Q_i \in \left\{ \xi_i^{(G)}, \xi_j^{(G)}\right\}$. Combined with Table 1, we can see that $U_i$ could be positive or it could be negative and $Q_i$ is all positive. Additionally, it is easy for us to see that $\alpha_i$ always has the same sign as $U_i$. Therefore, we can rewrite ($\textbf{P2-1-1}$) as follows with the condition  $g_i(\bm\Phi)>0$.
\begin{align}
\left( \textbf{P2-1-2} \right):&\min_{\bm\Phi} ~~~~~\sum_{i=1}^{2k_G}\alpha_i\\&  \nonumber
 ~\textrm{s.} \ \textrm{t.} ~~~~~~f_i(\bm\Phi)\leq0, ~~i=1,\cdots,2k_G\tag{34a}\\& \nonumber
~~~~~~~~~~\ |\bm{\Phi}_{n,n}|=1\tag{34b}
\end{align}
where
\begin{align}
f_i(\bm\Phi)&=h_i(\bm\Phi)-\alpha_ig_i(\bm\Phi)\nonumber\\&=
(U_i-\alpha_i)P_t\textrm{tr}(\textbf{D}_{ij}\bm\Phi^H\textbf{H}_{g}'^H\textbf{H}_{g}'\bm\Phi)-4\alpha_iQ_i\sigma_g^2
\end{align}
 Next, we aim to transform the constraints (34a) and (34b) to the convex constraints. First, we derive the first order and second order Hessian matrices of (34a) with respect to $\bm\Phi$ as follows
\begin{align}\label{first_order}
\nabla f_i(\bm\Phi)=(U_i-\alpha_i)P_t\textbf{H}^H\textbf{H}\bm\Phi\left(\textbf{D}_{ij}+\textbf{D}_{ij}^H\right)
\end{align}
\begin{align}\label{second_order}
\nabla^2f_i(\bm\Phi)=(U_i-\alpha_i)P_t\textbf{H}^H\textbf{H}\left(\textbf{D}_{ij}+\textbf{D}_{ij}^H\right)
\end{align}
which hold due to the fact that
\begin{align}
\frac{\partial\textrm{tr}(\textbf{WA}^H\textbf{W}^H\textbf{B})}{\partial\textbf{W}}=\textbf{BWA}+\textbf{B}^H\textbf{WA}^H
\end{align}
\begin{align}
\frac{\partial\textbf{AWB}}{\partial\textbf{B}}=\textbf{A}^H\textbf{B}
\end{align}

From (\ref{first_order}) and (\ref{second_order}) we can see that the convexity of constraint (34a) is determined by the positive and negative properties of $(U_i-\alpha_i)$, which is not sure as $U_i$ always has the same sign as $\alpha_i$. Notice that linear functions can be considered either convex or concave and we use SCA method to solve the above problem when $(U_i-\alpha_i)<0$ as follows
\begin{align}
f_i(\bm\Phi)\geq f_i(\bm\Phi_0)+\textrm{tr}\left[\nabla f_i(\bm\Phi_0)(\bm\Phi-\bm\Phi_0)\right]
\end{align}
Accordingly, the constraint (34a) can be rewritten as follows
\begin{equation}
\!\!f_i(\bm\Phi)\!\!=\!\!\!\left\{\!\!
\begin{aligned}
&\!(\!U_i\!-\!\alpha_i\!)\!P_t\textrm{tr}\!(\!\textbf{D}_{ij}\!\bm\Phi^H\!\textbf{H}_{g}'^H\!\textbf{H}_{g}'\!\bm\Phi)
\!-\!4\alpha_iQ_i\sigma_g^2,~~{(\!U_i\!-\!\alpha_i\!)\!\!>\!\!0}
\\&\!(\!U_i-\!\alpha_i\!)P_t\textrm{tr}(\textbf{D}_{ij}\!\bm\Phi_0^H\!\textbf{H}_{g}'^H\!\textbf{H}_{g}'\!\bm\Phi_0\!)
\!+\!(\!U_i\!-\!\alpha_i\!)\!\\&P_t\textrm{tr}\!\left(\!\textbf{H}^H\!\textbf{H}\bm\Phi\!\left(\textbf{D}_{ij}\!+\!\textbf{D}_{ij}^H\!\right)\left(\!\bm\Phi\!-\!\bm\Phi_0\right)\!\right)
-\!4\alpha_i,~~(\!U_i\!-\!\alpha_i\!)\!\!<\!\!0\\&\times Q_i\sigma_g^2
\end{aligned}
\right.
\end{equation}
And the constraint (34b) can be transformed to convex by relaxing (34b) as follows
\begin{align}
|\bm{\Phi}_{n,n}|\leq1
\end{align}
Hence, the optimization problem ($\textbf{P2-1-2}$) can be transformed to ($\textbf{P2-1-3}$) as follows
\begin{align}
\left( \textbf{P2-1-3} \right):&\min_{\bm\Phi} ~~~~~\sum_{i=1}^{2k_G}\alpha_i\\&  \nonumber
 ~\textrm{s.} \ \textrm{t.} ~~~~~~(41),(42)
\end{align}
which can be solved by using convex optimizing toolbox such as CVX. Then, we can get the beamformer at IRS as follows
\begin{align}
\bm{\Phi}_{n,n}=e^{j\textrm{angle}\left(\bm{\Phi}_{n,n}\right)},~~\forall n=1,2,\cdots,N.
\end{align}
Additionally, a step-by-step summary is provided as follows: $\textbf{Algorithm 1}$.
\begin{algorithm}
	\renewcommand{\algorithmicrequire}{\textbf{Input:}}
	\renewcommand{\algorithmicensure}{\textbf{Output:}}
	\caption{Proposed Max-NASR-SCA beamformer}
	\label{alg:1}
	\begin{algorithmic}[1]
		\REQUIRE the channel matrix $\textbf{H}_{B}'$ and $\textbf{H}_{E}'$, $P_t$, the $\mathcal{M}$-ary constellation
		\ENSURE $\bm\Phi$
		\STATE Initialize $\bm\Phi_0$ to a feasible value.
        \STATE Initialize $\alpha_i^0=\frac{h_i(\bm\Phi_0)}{g_i(\bm\Phi_0)}$, step $k=0$.
        \STATE Reformulate the problem by the SCA method to get $\textbf{P2-1-3}$
        \REPEAT
        \STATE Let $k=k+1$
        \STATE Update $\bm\alpha_{k}$ by (31)
        \STATE Update the beamforming matrix $\bm\Phi_{k}$ by solving the reformulated convex optimization problem (43)
        over $\bm\Phi_{k}$ for fixed $\bm\alpha_{k}$
        \UNTIL $\|\bm\Phi_{k}-\bm\Phi_{k-1}\|_2\leq0.01$
        \STATE Compute $\bm\Phi$ as the beamformer at IRS according to (44).
        \STATE \textbf{return} $\bm\Phi$
	\end{algorithmic}
\end{algorithm}
\subsection{Proposed Max-NASR-DA}
In the previous section, the Max-NASR-SCA algorithm was presented to optimize the secure IRS beamforming matrices. For the comparison of the secrecy performance and to offer a new solution to this non-convex optimization problem, we propose another secure IRS beamforming method with better performance, namely Max-NASR-DA, in what follows.

Observe the objective function below
\begin{align}
\left( \textbf{P2-1} \right):&\max_{\bm\Phi} ~\sum_{i=1}^{k_G}\frac{h_i(\bm\Phi)}{g_i(\bm\Phi)}
\\&\nonumber ~\textrm{s.} \ \textrm{t.} \ ~|\bm{\Phi}_{n,n}|=1,
\end{align}
it is easy to see that in each of the ratios, the denominator $g_i(\bm\Phi)>0, \forall i=1,2,\cdots,2k_G$, and the nominator $h_i(\bm\Phi)$ can be negative which
makes it difficult for us to realize quadratic transform. Therefore, we first realize the equivalent transform as follows
\begin{align}
\left( \textbf{P2-1-4} \right):&\max_{\bm\Phi} ~\sum_{i=1}^{2k_G}\frac{h_i(\bm\Phi)+M_i g_i(\bm\Phi)}{g_i(\bm\Phi)}-\sum_{i=1}^{2k_G}M_i
\\&\nonumber ~\textrm{s.} \ \textrm{t.} \ ~|\bm{\Phi}_{n,n}|=1,
\end{align}
where $M_i>\max |U_i|, i=1,2\cdots,2k_G$ and the nominator can be written as
\begin{align}
u_i(\bm\Phi)&=h_i(\bm\Phi)+M_ig_i(\bm\Phi)
\\&\nonumber=\underbrace{\left(U_i+M_i\right)P_t\textrm{tr}\left(\textbf{H}_{g}'\bm\Phi\textbf{D}_{ij}\bm\Phi^H\textbf{H}_{g}'^H\right)}_{a_1\geq0}+\underbrace{4M_iQ_i\sigma^2}_{a_2>0}.
\end{align}
According to (47), we can rearrange (46) as follows
\begin{align}
\left( \textbf{P2-1-4} \right):&\max_{\bm\Phi} ~\sum_{i=1}^{k_G}\frac{\left(U_i+M_i\right)P_t\textrm{tr}\left(\textbf{H}_{g}'\bm\Phi\textbf{D}_{ij}\bm\Phi^H\textbf{H}_{g}'^H\right)}{4Q_i\sigma_g^2+P_t\textrm{tr}\left(\textbf{H}_{g}'\bm\Phi\textbf{D}_{ij}\bm\Phi^H\textbf{H}_{g}'^H\right)}
\\&\nonumber~~+\!\!\sum_{i=1}^{2k_G}\!\frac{4M_iQ_i\sigma^2}{4Q_i\sigma_g^2+P_t\textrm{tr}\left(\textbf{H}_{g}'\bm\Phi\textbf{D}_{ij}\bm\Phi^H\textbf{H}_{g}'^H\right)}\!-\!\sum_{i=1}^{2k_G}\!M_i
\\&\nonumber ~\textrm{s.} \ \textrm{t.} \ ~|\bm{\Phi}_{n,n}|=1,
\end{align}
where each nominator and denominator are positive. Using quadratic transform proposed in \cite{shen20181,shen20182}, ($\textbf{P2-1-4}$) can be
reformulated as a equivalent optimization problem with a new objective function:
\begin{align}
f(\bm\Phi)=&\sum_{i=1}^{2k_G}2y_i\sqrt{\left(U_i+M_i\right)P_t}\textbf{H}_{g}'\bm\Phi\textbf{D}_{ij}^{-1/2}
\\&\nonumber+\sum_{i=1}^{2k_G}2y_i\sqrt{4M_iQ_i\sigma^2}
-\sum_{i=1}^{2k_G}\!M_i
\\&\nonumber-\sum_{i=1}^{2k_G}2y_i^2\left[4Q_i\sigma_g^2+P_t\textrm{tr}\left(\textbf{H}_{g}'\bm\Phi\textbf{D}_{ij}\bm\Phi^H\textbf{H}_{g}'^H\right)\right]
\end{align}
where the auxiliary variables $y_i$ can be found in a closed-form expression as follows
\begin{align}
y_i^*=\frac{\sqrt{u_i(\bm\Phi)}}{g_i(\bm\Phi)},~\forall i=1,\cdots,2k_G,
\end{align}
when $\bm\Phi$ is fixed.
($\textbf{P2-1-4}$) has been transformed as a non-convex optimization problem with convex objective function and
and non-convex equality constraint. According to the equality constraint, we adopt the dual ascent (DA) algorithm
to solve ($\textbf{P2-1-4}$).

The key idea of the non-convex DA is to introduce an auxiliary vector $\bm\varphi=\bm\Phi$, as well as a penalty term
for $\bm\varphi\neq\bm\Phi$. Then, ($\textbf{P2-1-4}$) is equivalently represented as
\begin{align}
\left( \textbf{P2-1-5} \right):&\max_{\bm\Phi} ~f(\bm\varphi)+\frac{\rho}{2}\|\bm\varphi-\bm\Phi\|_2^2
\\&\nonumber ~\textrm{s.} \ \textrm{t.} \ ~|\bm{\varphi}_{n,n}|=1,
\\&\nonumber~~~~~~~~\bm\varphi=\bm\Phi,
\end{align}
where $\rho>0$ is the penalty parameter. Then, we have the Lagrangian function of ($\textbf{P1.5}$):
\begin{align}
\mathcal G(\bm\varphi, \bm\Phi, \bm\lambda)=&\sum_{i=1}^{2k_G}2y_i^2\left[-4Q_i\sigma_g^2-P_t\textrm{tr}\left(\textbf{H}_{g}'\bm\varphi\textbf{D}_{ij}\bm\varphi^H\textbf{H}_{g}'^H\right)\right]
\\&\nonumber\!\!+\!\sum_{i=1}^{2k_G}2y_i\sqrt{\left(U_i+M_i\right)P_t}\textbf{H}_{g}'\bm\varphi\textbf{D}_{ij}^{-1/2}
\!-\!\!\sum_{i=1}^{2k_G}\!M_i
\\&\nonumber\!\!+\!\!\sum_{i=1}^{2k_G}\!2y_i\!\sqrt{4\!M_i\!Q_i\!\sigma^2}\!+\!\bm\lambda\!\left(\!\bm\varphi\!-\!\bm\Phi\!\right)\!+\!\frac{\rho}{2}\|\bm\varphi\!-\!\bm\Phi\|_2^2,
\end{align}
where $\bm\lambda$ is the dual variable for $\rm{R}\rm{e}$$\left\{\bm\varphi-\bm\Phi\right\}=0$ and
$\rm{I}\rm{m}$$\left\{\bm\varphi-\bm\Phi\right\}=0$, respectively. The alternating iterative process of DA including three main expressions:
\begin{align}
\bm\Phi^{t+1}=\textrm{arg} \max_{\bm\Phi} \mathcal G(\bm\varphi^t, \bm\Phi, \bm\lambda^t),
\end{align}
\begin{align}
\bm\varphi^{t+1}=\textrm{arg} \max_{\bm\varphi} \mathcal G(\bm\varphi, \bm\Phi^t, \bm\lambda^t),
\end{align}
\begin{align}
\bm\lambda^{t+1}=\bm\lambda^{t}+\rho(\bm\varphi^{t+1}-\bm\Phi^{t+1}),
\end{align}
where t is the iteration index.
\subsubsection{Optimizing $\bm\Phi$}
In (53), the optimal $\bm\Phi$ for fixed $\bm\varphi^t$ and $\bm\lambda^t$ is
\begin{align}
\bm\Phi^{t+1}=e^{j\textrm{angle}\left(\bm\varphi^t-\frac{1}{\rho}\bm\lambda^t\right)}.
\end{align}
\subsubsection{Optimizing $\bm\varphi$}
In (54), $\bm\varphi$ is optimized for fixed $\bm\Phi^t$ and $\bm\lambda^t$, and we have
\begin{align}
\bm\varphi^{t+1}&=\left[\rho\textbf{I}_{N}-\sum_{i=1}^{2k_G}2y_i^2P_t\textbf{H}_g'^H\textbf{H}_g'\left(\textbf{D}_{ij}+\textbf{D}_{ij}^H\right)\right]^{-1}
\\&\nonumber\times\!\!\left(\!-{\bm\lambda^t}^H\!-\!\sum_{i=1}^{2k_G}2y_i\sqrt{\left(U_i\!+\!M_i\right)P_t}\textbf{H}_{g}'\bm\varphi\textbf{D}_{ij}^{-1/2}\!+\!\rho\bm\Phi^{t+1}\right).
\end{align}
Additionally, a step-by-step summary is provided as follows: $\textbf{Algorithm 2}$.
\begin{algorithm}
	\renewcommand{\algorithmicrequire}{\textbf{Input:}}
	\renewcommand{\algorithmicensure}{\textbf{Output:}}
	\caption{Proposed Max-NASR-DA beamformer}
	\label{alg:1}
	\begin{algorithmic}[1]
		\REQUIRE the channel matrix $\textbf{H}_{B}'$ and $\textbf{H}_{E}'$, $P_t$, the $\mathcal{M}$-ary constellation
		\ENSURE $\bm\Phi$
		\STATE Initialize $\bm\Phi_0$, $\bm\varphi_0$ and $\bm\lambda_0$ to a feasible value.
        \STATE Initialize $y_i^0=\frac{h_i(\bm\Phi_0)}{g_i(\bm\Phi_0)}$, step $t=0$ and $\rho=0.5$.
        \STATE Reformulate the problem by the DA method to get $\textbf{P1.5}$
        \REPEAT
        \STATE Let $t=t+1$
        \STATE Update $y_{i}^t$ by (50)
        \REPEAT
        \STATE Update the beamforming matrix $\bm\Phi_{t+1}$ according to (56)
        \STATE Update the auxiliary matrix $\bm\varphi_{t+1}$ according to (57)
        \STATE Update the dual matrix $\bm\lambda_{t+1}$ according to (58)
        \UNTIL $\|\bm\Phi_{k}-\bm\Phi_{k-1}\|_2\leq0.01$
        \UNTIL $R_s(\bm\Phi_{t+1})-R_s(\bm\Phi_{t})\leq10^{-4}$
        \STATE \textbf{return} $\bm\Phi$
	\end{algorithmic}
\end{algorithm}
\subsection{Proposed Max-TASR-SDR method}
Based on the traditional approximate expression, we propose the Max-TASR-SDR method in this subsection. The expression of SR is given by
\begin{align}
R_s^a=I_0^B-I_0^E,
\end{align}
where $I_0^B$ and $I_0^E$ are shown in (11) and (12).
By using the Jensen's inequality and changing variables as $\bm{\Phi}\textbf{d}_{ij}=\rm{diag}\left\{ \textbf{d}_{ij} \right\}\bm{\varphi}$, the lower bound of $R_s^a$ is
\begin{align}\nonumber
&R_s^{a'}\\&\nonumber=\log_2\exp\sum_{i=1}^{GM}\sum_{j=1}^{GM}\left( \frac{-\bm{\varphi}^H\rm{diag}\left\{ \textbf{d}_{ij} \right\}^H\textbf{H}_E'^H\textbf{H}_E'\rm{diag}\left\{ \textbf{d}_{ij} \right\}\bm{\varphi}}{4} \right)
\\&-\log_2\exp\sum_{i=1}^{GM}\sum_{j=1}^{GM}\left( \frac{-\bm{\varphi}^H\rm{diag}\left\{ \textbf{d}_{ij} \right\}^H\textbf{H}_B'^H\textbf{H}_B'\rm{diag}\left\{ \textbf{d}_{ij} \right\}\bm{\varphi}}{4} \right)
\\&\nonumber=\log_2e\cdot\left[ \sum_{i=1}^{GM}\sum_{j=1}^{GM}\left( \frac{-\bm{\varphi}^H\rm{diag}\left\{ \textbf{d}_{ij} \right\}^H\textbf{H}_E'^H\textbf{H}_E'\rm{diag}\left\{ \textbf{d}_{ij} \right\}\bm{\varphi}}{4} \right) \right .
\\&~~~\left . -\sum_{i=1}^{GM}\sum_{j=1}^{GM}\left( \frac{-\bm{\varphi}^H\rm{diag}\left\{ \textbf{d}_{ij} \right\}^H\textbf{H}_E'^H\textbf{H}_E'\rm{diag}\left\{ \textbf{d}_{ij} \right\}\bm{\varphi}}{4} \right) \right].
\end{align}
Hence, Problem $\left( \textbf{P2-1-6} \right)$ can be rewritten as follows
\begin{align}\label{p1.1}
\left( \textbf{P2-1-6} \right):~&\max ~\frac{\log_2e}{4}\cdot\bm{\varphi}^H\bm{\Omega}\bm{\varphi}\\&\nonumber \textrm{s.} \ \textrm{t.} \ ~~|\bm{\varphi_{n}}|=1,
\end{align}
where
\begin{align}\nonumber
\bm{\Omega}=&\sum_{i=1}^{GM}\sum_{j=1}^{GM}\left( -\rm{diag}\left\{ \textbf{d}_{ij} \right\}^H\textbf{H}_E'^H\textbf{H}_E'\rm{diag}\left\{ \textbf{d}_{ij} \right\} \right)
\\&-\sum_{i=1}^{GM}\sum_{j=1}^{GM}\left( -\rm{diag}\left\{ \textbf{d}_{ij} \right\}^H\textbf{H}_B'^H\textbf{H}_B'\rm{diag}\left\{ \textbf{d}_{ij} \right\} \right).
\end{align}

The objective function of (\ref{p1.1}) can be rewritten as $\bm{\varphi^H\Omega\varphi}=\rm{tr} \left( \bm{\Omega}\textbf{Q} \right)$ with $\textbf{Q}=\bm{\varphi\varphi^H}$. In particular, $\textbf{Q}$ is a positive semidefine matrix with $ \rm{rank}(\textbf{Q})=1$. However, as the rank-one constraint is non-convex, we apply the semi-definite relaxation (SDR) method to relax this constraint and reformulate (\textbf{P2-1-6}) as
\begin{align}\label{p1.2}
\left( \textbf{P2-1-7} \right):~&\max ~\frac{\log_2e}{4}\cdot\rm{tr}( \bm{\Omega}\textbf{Q} )\\&\nonumber \textrm{s.} \ \textrm{t.} \ ~~\textbf{Q}_{n,n}=1,\\&\nonumber~~~~~~~\textbf{Q}\succeq0,
\end{align}
which is a standard convex semi-definite programming (SDP) problem and can be solved via exsiting convex optimization solvers such as CVX \cite{MGCVX}. It is worth pointing out that after the relaxation, the optimal solution $\textbf{Q}^*$ to problem (\textbf{P2-1-7}) may not be a rank-one solution and we can solve it by the method of Gaussian randomization.

\section{TRANSMIT POWER DESIGN FOR GIVEN BEAMFORMING BASED ON APPROXIMATE
EXPRESSION OF SR}
In IRS-aided SM systems, transmit power design is an important technique to improve the system performance. In
this section, two transmit power design methods are proposed based on the TASR and proposed NASR expression
respectively, for secure IRS-aided SM systems: Max-NASR-PTD amd Max-TASR-PTD.
\subsection{Transmit Power Design based on Proposed NASR}
As the beamformer is fixed, we can write the optimization problem as follows
\begin{align}
\left( \textbf{P2-2} \right):&\max_{\beta} ~~~~~~\sum_{i=1}^{2k_G}\frac{h_i\left(\beta\right)}{g_i\left(\beta\right)}
\\&\nonumber \textrm{subject} \ \textrm{to} \ \beta^2\leq1
\end{align}
where
\begin{align}
h_i(\beta)=U_i \beta^2\textrm{tr}(\textbf{D}_{ij}\bm\Phi^H\textbf{H}_{g}'^H\textbf{H}_{g}'\bm\Phi),
\end{align}
and
\begin{align}
g_i(\beta)=4Q_i\sigma_g^2+\beta^2\textrm{tr}(\textbf{D}_{ij}\bm\Phi^H\textbf{H}_{g}'^H\textbf{H}_{g}'\bm\Phi),
\end{align}
where g stands for B (Bob) or E (Eve), $U_i \in \left\{ \zeta_i^{(G)}, -\zeta_j^{(G)}\right\}$, $Q_i \in \left\{ \xi_i^{(G)}, \xi_j^{(G)}\right\}$. Similarly, to ensure the denominator and the nominator are all positive, we transform the optimization problem
as follows equally
\begin{align}
\left( \textbf{P2-2-1} \right):&\max_{\beta} ~~~~~~\sum_{i=1}^{2k_G}\frac{h_i(\beta)+M_i g_i(\beta)}{g_i(\beta)}-\sum_{i=1}^{2k_G}M_i\\&\nonumber \textrm{subject} \ \textrm{to} \ \beta^2\leq1,
\end{align}
where $M_i>\max|U_i|,i=1,2,\cdots,2k_G$ and therefore the nominator and the denominator in each ratio are positive. Similarly, we can
rearrange (67) as follows
\begin{align}
\left( \textbf{P2-2-1} \right):&\max_{\beta} ~\sum_{i=1}^{k_G}\frac{\left(U_i+M_i\right)\beta^2P_s\textrm{tr}\left(\textbf{H}_{g}'\bm\Phi\textbf{D}_{ij}\bm\Phi^H\textbf{H}_{g}'^H\right)}{4Q_i\sigma_g^2+\beta^2P_s\textrm{tr}\left(\textbf{H}_{g}'\bm\Phi\textbf{D}_{ij}\bm\Phi^H\textbf{H}_{g}'^H\right)}
\\&\nonumber~~+\!\!\sum_{i=1}^{2k_G}\!\frac{4M_iQ_i\sigma^2}{4Q_i\sigma_g^2+\beta^2P_s\textrm{tr}\left(\textbf{H}_{g}'\bm\Phi\textbf{D}_{ij}\bm\Phi^H\textbf{H}_{g}'^H\right)}\!-\!\sum_{i=1}^{2k_G}\!M_i
\\&\nonumber ~\textrm{s.} \ \textrm{t.} \ ~\beta^2\leq1,
\end{align}

Using quadratic transform proposed in \cite{shen20181}, ($\textbf{P2}$) can be
reformulated as an equivalent optimization problem with a new objective function:
\begin{align}
f(\beta)=&\sum_{i=1}^{2k_G}2y_i\sqrt{\left(U_i+M_i\right)\textrm{tr}\left(\textbf{H}_{g}'\bm\Phi\textbf{D}_{ij}\bm\Phi^H\textbf{H}_{g}'^H\right)}\beta^2P_s
\\&\nonumber+\sum_{i=1}^{2k_G}2y_i\sqrt{4M_iQ_i\sigma^2}
-\sum_{i=1}^{2k_G}\!M_i
\\&\nonumber-\sum_{i=1}^{2k_G}2y_i^2\left[4Q_i\sigma_g^2+\beta^2P_s\textrm{tr}\left(\textbf{H}_{g}'\bm\Phi\textbf{D}_{ij}\bm\Phi^H\textbf{H}_{g}'^H\right)\right]
\end{align}
where the auxiliary variables $y_i$ can be found in a closed-form expression as follows
\begin{align}
y_i^*=\frac{\sqrt{h_i(\beta)+M_i g_i(\beta)}}{g_i(\beta)},~\forall i=1,\cdots,2k_G,
\end{align}
when $P_t=\beta^2 P_s$ is fixed.

Then, we have the Lagrangian function of ($\textbf{P2-2-1}$):
\begin{align}
\mathcal L\left( \beta, \lambda \right)=&\sum_{i=1}^{2k_G}2y_i\sqrt{\left(U_i+M_i\right)\textrm{tr}\left(\textbf{H}_{g}'\bm\Phi\textbf{D}_{ij}\bm\Phi^H\textbf{H}_{g}'^H\right)}\beta
\\&\nonumber+\sum_{i=1}^{2k_G}2y_i\sqrt{4M_iQ_i\sigma^2}
-\sum_{i=1}^{2k_G}\!M_i
-\lambda\left(\beta^2P_s-1\right)
\\&\nonumber-\sum_{i=1}^{2k_G}2y_i^2\left[4Q_i\sigma_g^2+\beta^2P_s\textrm{tr}\left(\textbf{H}_{g}'\bm\Phi\textbf{D}_{ij}\bm\Phi^H\textbf{H}_{g}'^H\right)\right].
\end{align}
And we get the KKT condition equations as follows
\begin{equation}
\left\{
\begin{aligned}
&\nabla_{\beta}\mathcal L\left( \beta, \lambda \right)=0,
\\&\beta^2P_s-1\geq0,
\\&\lambda(\beta^2P_s-1)=0,
\\&\lambda\geq0.
\end{aligned}
\right.
\end{equation}
which yields the solution
\begin{equation}
\left\{
\begin{aligned}
&\beta=\frac{\sum_{i=1}^{2k_G}y_i\sqrt{\left(U_i+M_i\right)P_s\textrm{tr}\left(\textbf{H}_{g}'\bm\Phi\textbf{D}_{ij}\bm\Phi^H\textbf{H}_{g}'^H\right)}}{\lambda+\sum_{i=1}^{2k_G}y_i^2P_s\textrm{tr}\left(\textbf{H}_{g}'\bm\Phi\textbf{D}_{ij}\bm\Phi^H\textbf{H}_{g}'^H\right)},
\\&\lambda=0.
\end{aligned}
\right.
\end{equation}
Additionally, a step-by-step summary is provided as follows: $\textbf{Algorithm 3}$
\begin{algorithm}
	\renewcommand{\algorithmicrequire}{\textbf{Input:}}
	\renewcommand{\algorithmicensure}{\textbf{Output:}}
	\caption{Proposed Max-NASR-PTD transmit power method}
	\label{alg:1}
	\begin{algorithmic}[1]
		\REQUIRE the channel matrix $\textbf{H}_{B}'$ and $\textbf{H}_{E}'$, $\bm\Phi$, $P_s$, the $\mathcal M$-ary constellation
		\ENSURE $\beta$
		\STATE Initialize $\beta^0$ to a feasible value.
        \STATE Initialize $y_i^0$ according to (70), step $k=0$.
        \STATE Reformulate the problem by the quadratic transform to get (68).
        \REPEAT
        \STATE Update $k=k+1$
        \STATE Update $y_i^k$ according to (70)
        \STATE Update $\beta^k$ by (73)
        \UNTIL{$R_s(\beta^k)-R_s(\beta^{k-1})\leq10^{-4}$}

        \STATE \textbf{return} $P_t$
	\end{algorithmic}
\end{algorithm}
\subsection{Transmit Power Design based on TASR}
In the previous section, the Max-NASR-PTD algorithm was presented to optimize the transmit power
for higher SR. For the comparison of  the secrecy performance and  to offer a new solution to
this non-convex optimization problem, we propose another secure transmit power scheme,
namely Max-TASR-PTD based on the TASR expression, in what follows.
To maximize $R_s^a(P_t)$, the Max-TASR-PTD method can be employed to directly optimize
the transmit power $P_t$. We derive the gradient of $R_s^a(P_t)$  with respect to $P_t$ applying $P_t=\beta^2 P_s$ as
\begin{equation}
\begin{aligned} \label{gradient}
&\!\nabla_{\beta} R_s^{a}(\beta)=\frac{- \beta \sqrt{P_s}}{2\sigma^2\ln2}\times\!\!\left[\!\!\frac{1}{\kappa_E}\!\sum \limits_{i=1}^{GM}\!\sum \limits_{j=1}^{GM}\!\exp\!\!\left(\beta^2P_s\alpha_{i,j}^e\right)\cdot\alpha_{i,j}^e\right.\\&\left.-\frac{1}{\kappa_B}\!\sum \limits_{i=1}^{GM}\!\sum \limits_{j=1}^{GM}\!\exp\!\!\left(\beta^2P_s\alpha_{i,j}^b\right)\cdot\alpha_{i,j}^b\right],
\end{aligned}
\end{equation}
where
\begin{align}
\alpha_{i,j}^b=-\frac{\textbf{d}_{i,j}^H\textbf{H}_{B}'^H\bm\Phi^H\bm\Phi\textbf{H}_{B}\textbf{d}_{i,j}}{4\sigma^2},
\end{align}
\begin{align}
\alpha_{i,j}^e=-\frac{\textbf{d}_{i,j}^H\textbf{H}_{E}'^H\bm\Phi^H\bm\Phi\textbf{H}_{E}\textbf{d}_{i,j}}{4\sigma^2},
\end{align}
\begin{align}
\kappa_B=\log_2\sum_i^{GM}\sum_j^{GM}\exp\left(\beta^2P_s\alpha_{i,j}^B\right),
\end{align}
\begin{align}
\kappa_E=\log_2\sum_i^{GM}\sum_j^{GM}\exp\left(\beta^2P_s\alpha_{i,j}^E\right).
\end{align}
In order to find a locally optimal  $\beta$, we first initialize $\beta$ and $R_s^{a}$,  solve the gradient
$\nabla_{\beta} R_s^{a}(\beta)$, and adjust $P_t$ according to $\nabla_{\beta} R_s^{a}(\beta)$. The value of $\beta$
 is updated according to the following iterative formula
\begin{align}
\beta_{k+1}=\beta_{k}+\mu\nabla_{\beta} R_s^{a}(\beta_k).
\end{align}
Then, obtain $R_s^{a}$, update $\beta$ or step size $\mu$ according to the
difference between before and after $R_s^{a}$, and repeat the above
steps until the termination condition is reached.
\section{Simulation Results and Analysis}
In this section, we evaluate the performance of these beamformers and these transmit power methods.  The system parameters are set as follows: $G=4$, $N=100$, $N_t=1$, $P_s=N_t $W and quadrature phase shift keying (QPSK) modulation\cite{zhu2009chunk,zhu2011chunk}. For the convenience of simulation, it is assumed that the total transmit power $P=N_t$ W. For the sake of fairness of Bob and Eve, it is assumed that all noise variances in  channels are identical, i.e., $\sigma_b^2=\sigma_e^2$ and  $N_b=N_e=2$.

We compare our proposed algorithms to the following benchmark schemes:

(1)~Case I:~\textbf{IRS with no beamforming:} Obtain the maximum SR by optimizing beamforming vectors with the IRS phase-shift matrix set to zero with the magnitude set to one, i.e., $\bm\Phi=\textbf{I}_{N\times N}$.

(2)~Case II:~\textbf{IRS with Random beamforming:} Obtain the maximum SR by optimizing the beamforming vectors with all the phase for each reflection element uniformly and independently generated from [0,2$\pi$).
\begin{figure}
  \centering
  \includegraphics[width=0.42\textwidth]{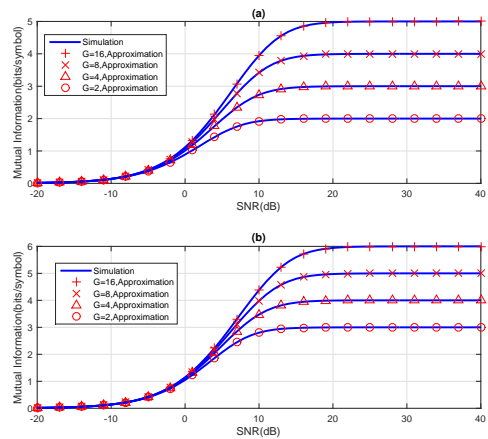}\\
  \caption{Simulated (denoted by lines) and approximated mutual information
of M-PSK (a.BPSK, b.QPSK) over Rayleigh fading channels}
\label{fixSR}
\end{figure}

Fig.~\ref{fixSR} demonstrates the mutual information fitting results for the NASR which is proposed in Section-II
with BPSK and QPSK employed. For BPSK (Fig.~2(a)), the simulation results of NASR fits the curves of practical simulation
results well with different number of IRS subsets in all SNR regions. As for QPSK (Fig.~2(b)), we have the same the
conclusion about the accuracy of fitness.
\begin{figure}[ht]
\centerline{\includegraphics[width=0.42\textwidth]{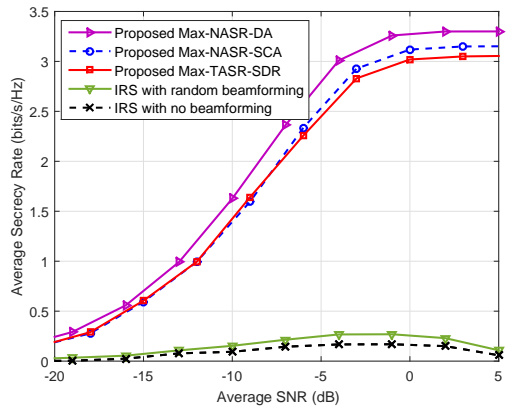}}
\caption{Curves of average SR versus average SNR for different IRS beamforming algorithms with transmit power fixed with $G=4$.}
\label{changeBf}
\end{figure}

When the  Max-NASR-TPD transmit power strategy is adopted in IRS-aided SM, Fig.~\ref{changeBf} demonstrates the
curves of average SR versus SNR of the proposed Max-NASR-DA, proposed Max-NASR-SCA and proposed Max-TASR-SDR for $G=4$
with the the above Case I and Case II as the performance benchmarks. It
is seen that the SR of the proposed Max-NASR-DA algorithm is much better than that of Max-NASR-SCA and Max-TASR-SDR
in all average SNR regions.  As the average SNR increases, the Max-NASR-DA algorithm can approach the rate ceil of the achievable SR with a more rapid rate than  Max-NASR-SCA and Max-TASR-SDR. This tendency implies that
the SR performance of the proposed Max-NASR-DA beamformer is better than that of  Max-TASR-SDR and Max-NASR-SCA.
The SR  performance of the proposed Max-NASR-SCA method is in a region between the proposed Max-TASR-SDR and Max-NASR-DA
in the medium and high average SNR regions, and only slightly higher than Max-TASR-SDR algorithm in the low
average SNR region.
\begin{figure}[ht]
\centerline{\includegraphics[width=0.42\textwidth]{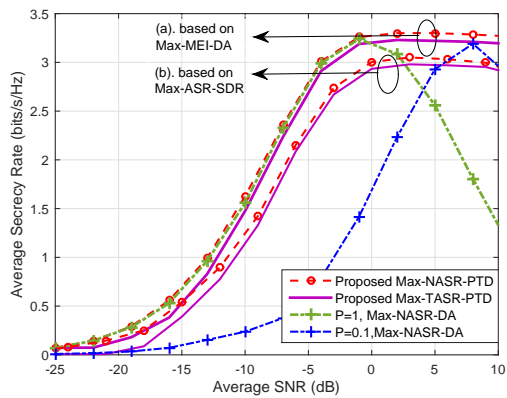}}
\caption{Curves of average SR versus average SNR for different TPD methods with IRS beamformer fixed with $G=4$.}
\label{changePt}
\end{figure}

Now, we fix the IRS beamforming method, and make a performance comparison of two PTD strategies. Fig.~\ref{changePt} demonstrates
the average SR versus average SNR of two PTD methods when $G=4$ and $N=100$. Fig.~4(a), we adopt
the Max-NASR-DA method as IRS beamformer, it can be clearly seen that the proposed Max-NASR-PTD strategy has higher security
performance than Max-TASR-PTD nearly in all average SNR regions. Meanwhile, Fig.~4(b) which adopt
Max-TASR-SDR method as the fixed IRS beamformer, the tendency of the curves is the same as those in Fig.~4(a). However,
in all SNR regions, the proposed two methods outperforms two fixed transmit power strategies in terms of SR.
This confirms that PTD can improve the SR performance. Observing Fig.~\ref{changePt}, we find the SRs of Max-NASR-PTD and
Max-TASR-PTD increase as SNR increases. For the three fixed PTD strategies, their SRs first increase up to the
corresponding maximum values, and then reduce gradually as SNR increases. The main reason is that their transmit power
are fixed and independent of the change of SNR while the remaining two schemes adaptively adjust their transmit power
in accordance with the exact value of SNR in channel.
\begin{figure}[ht]
\centerline{\includegraphics[width=0.42\textwidth]{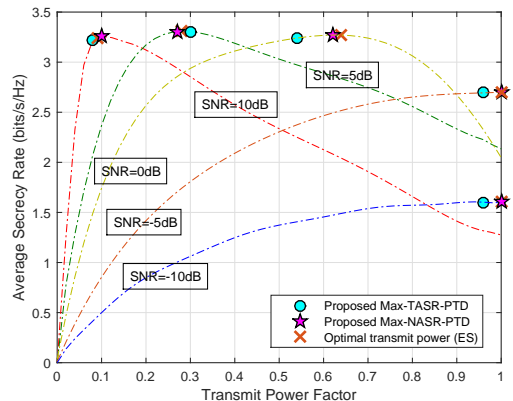}}
\caption{Curves of average SR versus average $\beta$ for different IRS beamforming algorithms with different SNR and $G=4$.}
\label{Pt}
\end{figure}

Fig.~\ref{Pt} plots the maximum achievable SR versus transmit power with different SNR. From this figure,
we can see that for all fixed SNR, the Max-NASR-PTD always approaches the value got from optimal exhaustive search (ES)
better, where the configurations are the same as Fig.~\ref{changePt}. Additionally, it can be observed that when
SNR=0dB, the change of SR versus transmit power is generally gentle, while there is a slightly big difference
between the transmit power solved by different algorithms, but the corresponding SR value does not change much.
As for under the other values of SNR, it can be observed that the change of SR versus transmit power becomes steeper,
and the solutions obtained by the three different algorithms are relatively close.
\begin{figure}[ht]
\centerline{\includegraphics[width=0.42\textwidth]{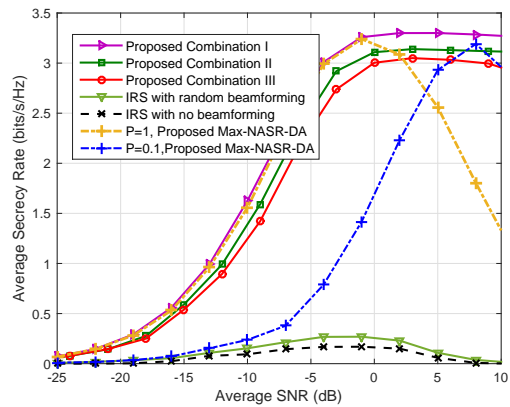}}
\caption{Curves of average SR versus average SNR for different IRS beamforming algorithms and TPD algorithms with $G=4$.}
\label{PtPlusBF}
\end{figure}

Fig.~\ref{PtPlusBF} plots the curves of the average SR versus average SNR for jointly optimizing IRS beamformer and transmit
power with the previous Case I and Case II as the performance benchmarks. There are three combinations for the beamformer and
PTD: 1).Combination I:Max-NASR-DA plus Max-NASR-PTD; 2).Combination II:Max-NASR-SCA plus Max-NASR-PTD;
3).Combination III:Max-TASR-SDR plus Max-TASR-PTD. From Fig.~\ref{PtPlusBF}, it is obviously seen that
the SR performance of Combination I performs much better than those of other combinations as Max-NASR-DA and
Max-NASR-PTD both have the best security performance. Compared with Combination III, Combination I and Combination II
harvest more SR performance gains in all SNR regions. And Combination II is between the proposed Combination I
and Combination III in most SNR regions.

\begin{figure}[ht]
\centerline{\includegraphics[width=0.42\textwidth]{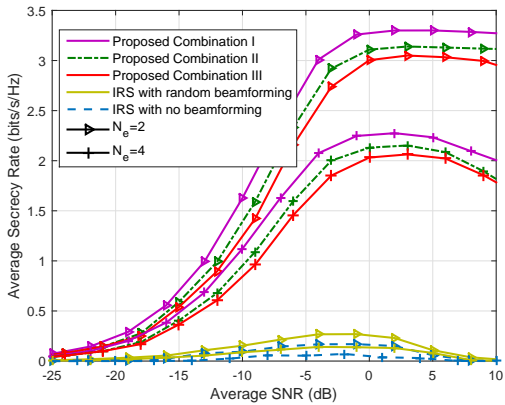}}
\caption{Curves of average SR versus average SNR for different IRS beamforming algorithms and TPD algorithms with $G=4$ and different $N_e$.}
\label{changeNe}
\end{figure}
Fig.~\ref{changeNe} plots the average SR performance achieved by the above three combinations
by fixing $G\!=\!4$ and $N_b\!=\!2$, and only changing $N_e$.
When $N_b\!=$ $N_e\!=\!2$, as average SNR increases, the SR curves increase until it reaches a certain average SNR,
and then the curves will descend slightly. However, when $N_e$ is larger than $N_b$, the
SR will descend much when the average SNR is beyond some thresholds.  As $N_e$ increases, the SR curves can achieve
the corresponding maximum SRs, and then  decreases. This is because when average SNR is very high, both Bob and Eve
have very good quality of channels, while Eve has a larger number of receive antennas than Bob, which is the worst
situation, so the SR performance begins to decline.
\begin{figure}[ht]
\centerline{\includegraphics[width=0.42\textwidth]{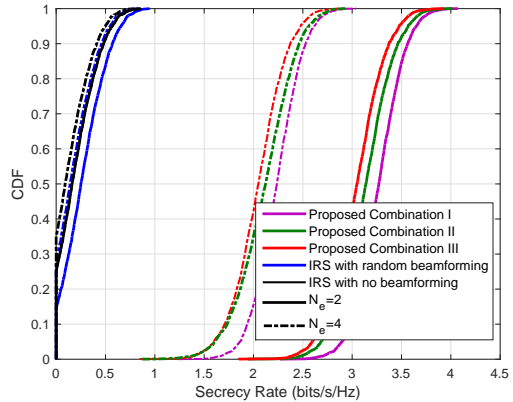}}
\caption{CDF curves versus SR with $G=4$ and different $N_e$.}
\label{cdf}
\end{figure}

Fig.~\ref{cdf} shows the cumulative distribution function (CDF) curves of the three combinations for the different
numbers of eavesdropper's antennas when average SNR = 5dB. In this situation, Fig.~\ref{cdf} has
the same descending trend in SR performance as Fig.~\ref{changeNe}: Combination I, Combination II and Combination III.

\begin{figure}[ht]
\centerline{\includegraphics[width=0.42\textwidth]{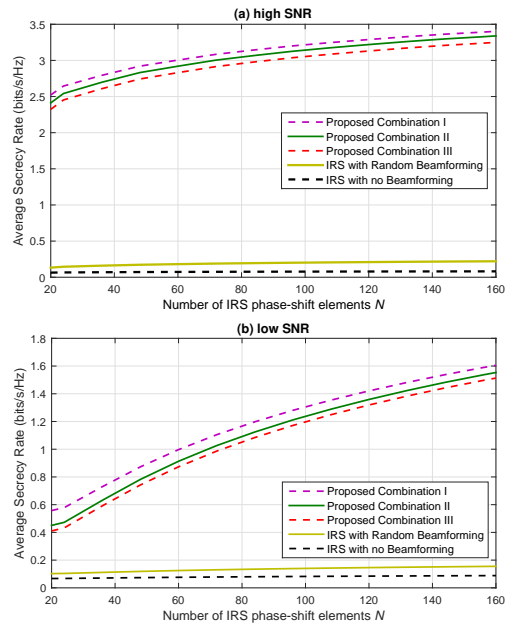}}
\caption{Curves of average SR versus number of IRS phase-shift elements for different IRS beamforming algorithms and PTD algorithms with $G=4$ and different $N_e$.}
\label{M}
\end{figure}

In Fig.~\ref{M}, we investigate the impact of the numbers of IRS phase-shift elements for SR performance under
(a).low-SNR region and (b).high-SNR region, respectively. From Fig.~\ref{M} (a) and (b), it can be seen that the
proposed three combinations improve the SR performance whether in the low-SNR regime or the high-SNR regime.
As the number of IRS elements increases, the SR gains achieved by Combination I and Combination II over IRS with
no beamforming and random phase grow gradually and become more significant. Compared with IRS with
no beamforming and random phase scheme, the IRS phase-shift-optimization schemes and transmit power design
performs much better, especially with a large value of $N$. This is explicit about the importance of the optimization
of the phase-shift design. Even with a value of $N=30$, our proposed scheme can also perform better than that
scheme without the IRS phase-shift-optimization.
\section{Conclusion}
In this paper, we have made a comprehensive investigation of IRS beamforming and transmit power design concerning
IRS-SSM. In such an architecture, the first part of bitstream is transmitted by APM symbol, and the second part
of bitstream is carried by selecting a subset in the IRS rather than a single transmit/receiver antenna. Considering
the physical-layer security, a simple approximated SR expression was proposed. Based on the NASR and TASR, three
IRS beamformers, Max-NASR-DA, Max-NASR-SCA and Max-TASR-SDR, were proposed. Simulation results showed that the proposed
beamforming methods have an ascending order in SR: IRS with no beamforming, IRS with random beamforming, Max-TASR-SDR, Max-NASR-SCA and Max-NASR-DA. Particularly, two PTD methods were also proposed: Max-NASR-PTD and Max-TASR-PTD. Simulation results showed that the proposed PTD strategies have an ascending order in SR: fixed transmit power, Max-TASR-PTD and Max-NASR-PTD. Accordingly, either the IRS beamforming or the PTD algorithms based on the NASR performs better than those based on the TASR.

\vspace{12pt}

\bibliographystyle{IEEEtran}
\bibliography{IEEEabrv,refx}
\end{document}